\documentclass[11pt]{article} 
\usepackage[margin=1in]{geometry} 

\usepackage{natbib}
\bibpunct{(}{)}{;}{a}{}{,} 
\usepackage{xcolor} 
\definecolor{darkblue}{HTML}{090F7A}

\usepackage{hyperref}
\hypersetup{
    colorlinks=true,        
    breaklinks=true,        
    linkcolor=darkblue,     
    urlcolor=darkblue,      
    anchorcolor=darkblue,   
    citecolor=darkblue,     
}

\usepackage{scalerel,graphicx}
\definecolor{orcidlogocol}{HTML}{A6CE39}
\newcommand{\orcidicon}{\includegraphics[width=8pt]{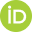}}
\newcommand{\orcid}[1]{\href{https://orcid.org/#1}{\textcolor{orcidlogocol}{\orcidicon}}}

\title{\texttt{pyROX}: Rapid Opacity X-sections}
\author{
    Sam de Regt\,\orcid{0000-0003-4760-6168}\textsuperscript{1}, 
    Siddharth Gandhi\,\orcid{0000-0001-9552-3709}\textsuperscript{2,3}, 
    Louis Siebenaler\,\orcid{0009-0005-3389-8819}\textsuperscript{1}, 
    Dar\'io Gonz\'alez Picos\,\orcid{0000-0001-9282-9462}\textsuperscript{1}
}

\date{}

\begin{document}
\maketitle

\begin{flushleft}
    \textsuperscript{1} Leiden Observatory, Leiden University, P.O. Box 9513, 2300 RA, Leiden, The Netherlands\\
    \textsuperscript{2} Department of Physics, University of Warwick, Coventry CV4 7AL, UK\\
    \textsuperscript{3} Centre for Exoplanets and Habitability, University of Warwick, Gibbet Hill Road, Coventry CV4 7AL, UK
\end{flushleft}

\section*{Summary}
In recent years, significant advances have been made in exoplanet and brown dwarf observations. By using state-of-the-art models, astronomers can determine properties of their atmospheres, such as temperatures, the presence of clouds, or the chemical abundances of molecules and atoms. Accurate and up-to-date opacities are crucial to avoid inconclusive or biased results, but it can be challenging to compute opacity cross-sections from the line lists provided by various online databases. 

We introduce \texttt{pyROX}, an easy-to-use Python package to calculate molecular and atomic cross-sections. Since \texttt{pyROX} works on CPUs, it can compute a small line list on a regular workstation, but it is also easily parallelised on a cluster for larger line lists. In addition to line opacities, \texttt{pyROX} also supports calculations of collision-induced absorption. Tutorials are provided in the online documentation which explain the configuration parameters and different functionalities of \texttt{pyROX}.

\section*{Statement of need}
The advent of a new generation of telescopes and instruments has led to a dramatically increased quality in observations of exoplanets and brown dwarfs. Such sub-stellar objects are now observed over a wide wavelength range (1-20 µm) with JWST spectra (e.g. \citealt{August_ea_2023, Carter_ea_2024, Matthews_ea_2025, Miles_ea_2023}), for instance, which was previously difficult to access. Developments in ground-based instrumentation allow astronomers to measure young exoplanet companions at closer separations to their host stars (e.g. \citealt{Landman_ea_2024, Xuan_ea_2024b}) and at high spectral resolutions (e.g. \citealt{Nortmann_ea_2025, Xuan_ea_2024a}). At the same time, progress has also been made in atmospheric modelling using software for radiative transfer, chemistry, circulation models, etc. (e.g. \citealt{Molliere_ea_2019, Stock_ea_2018, Wardenier_ea_2021}). Recently, these observations and software are coupled with sampling algorithms to characterise the atmospheres of the sub-stellar objects (e.g. \citealt{Barrado_ea_2023, Brogi_ea_2019, Gibson_ea_2020, Line_ea_2015}). 

Opacity cross-sections play a key role in accurately modelling sub-stellar atmospheres. Opacity governs the dominant energy transport mechanism (i.e. radiation or convection) which affects the thermal structure of the atmosphere \citep{Marley_ea_2021}. Furthermore, high-resolution studies require well-determined frequencies for the transition lines. Inaccuracies in line-list data can result in biased abundance constraints (e.g. \citealt{Brogi_ea_2019, de_Regt_ea_2024}) or ambiguous (non)-detections of certain molecules (e.g. \citealt{de_Regt_ea_2022, Merritt_ea_2020, Serindag_ea_2021}). It is therefore important that the most up-to-date and complete opacity data are used. However, it can be difficult to efficiently calculate opacity cross-sections from line lists that sometimes consist of billions of transitions. 

To help resolve this challenge, we present \texttt{pyROX}, a user-friendly Python package to calculate molecular and atomic cross-sections for applications in models of sub-stellar atmospheres. \texttt{pyROX} supports line opacity calculations from the ExoMol \citep{Tennyson_ea_2024}, HITRAN \citep{Gordon_ea_2022}, HITEMP \citep{Rothman_ea_2010}, and Kurucz\footnote{\url{http://kurucz.harvard.edu/}} databases. Collision-Induced Absorption (CIA) coefficients can also be calculated from the HITRAN and Borysow\footnote{\url{https://www.astro.ku.dk/~aborysow/programs/index.html}} databases. So far, \texttt{pyROX}-computed cross-sections have enabled several recent publications from our research group \citep{de_Regt_ea_2025, Siebenaler_ea_2025}.

\section*{Functionality of \texttt{pyROX}}
Documentation for \texttt{pyROX} is available at \url{https://py-rox.readthedocs.io/en/latest/} and includes tutorial examples to running the code. Here, we outline the main functionality of \texttt{pyROX}:

\begin{itemize}
    \item[-] \textbf{Download and read files}: The necessary input files (line lists, partition functions, broadening coefficients or CIA files) can be downloaded with a simple command. When reading the relevant parameters, \texttt{pyROX} handles the different data structures of the supported databases (ExoMol, HITRAN/HITEMP, Kurucz, Borysow). 
    \item[-] \textbf{Compute line-strengths and -widths}: For line-opacity calculations, \texttt{pyROX} calculates the strength and broadening-widths for each line transition at the user-provided pressure and temperature. Support is offered for various \href{https://py-rox.readthedocs.io/en/latest/notebooks/pressure_broadening.html}{pressure-broadening descriptions}. \texttt{pyROX} can speed up the line-profile computation by selecting only the main line-strength contributors.
    \item[-] \textbf{Compute line profiles}: Next, \texttt{pyROX} computes the Voigt profiles as the real part of the Faddeeva function (Eq. 12 of \citealt{Gandhi_ea_2020}), using the \texttt{scipy.special.wofz} implementation\footnote{\url{https://docs.scipy.org/doc/scipy/reference/generated/scipy.special.wofz.html}}. 
    \item[-] \textbf{Combine and save}: The line profiles are summed into wavelength-dependent cross-sections for each temperature-pressure point. These cross-sections are saved into an efficient HDF5 output file. For CIA calculations, \texttt{pyROX} restructures the coefficients read from the input files into a wavelength- and temperature-dependent grid and also saves these data to an HDF5 file.
\end{itemize}

Currently, \texttt{pyROX} offers built-in support for converting its output into the high-resolution opacities used by \texttt{petitRADTRANS} \citep{Molliere_ea_2019}. In future releases, we plan to add conversions for other radiative transfer codes that are popular in the exoplanet and brown dwarf community. We welcome suggestions for new features, which can be done by \href{https://github.com/samderegt/pyROX/issues}{opening an issue} on GitHub. If you want to contribute to \texttt{pyROX}, please read the \href{https://py-rox.readthedocs.io/en/latest/contributing.html}{documented guidelines}.

\section*{Similar tools}
Existing open source codes, such as \href{https://github.com/MartianColonist/Cthulhu}{\texttt{Cthulhu}} \citep{Agrawal_ea_2024}, \href{https://github.com/Trovemaster/exocross}{\texttt{ExoCross}} \citep{Yurchenko_ea_2018, Zhang_ea_2024} and \href{https://github.com/exoclime/HELIOS-K}{\texttt{HELIOS-K}} \citep{Grimm_ea_2015, Grimm_ea_2021}, can calculate cross-sections at comparable performances to \texttt{pyROX}. However, \texttt{ExoCross} is written in Fortran and \texttt{HELIOS-K} utilises GPU-acceleration which can limit their use to experts with the appropriate hardware. \texttt{pyROX} is a Python code that runs only on CPUs which should make it accessible for the opacity needs of most astronomers. Notably, \texttt{pyROX} supports cross-section calculations on any user-provided wavelength or wavenumber grid. This enables the user to fix the spectral resolution ($\mathcal{R}=\lambda/\Delta\lambda$) which cannot be achieved with equal wavelength- or wavenumber-spacing.

\section*{Acknowledgements}
S.d.R. acknowledges funding from NWO grant OCENW.M.21.010.

{
\bibliography{refs.bib}
}

\end{document}